\documentclass[twocolumn,aps,prx,superscriptaddress]{revtex4-1}
\usepackage{graphicx}

\begin{document}

\title{Destroying coherence in high temperature superconductors with current flow}

\author{A. Kaminski}
\affiliation{Ames Laboratory and Department of Physics and Astronomy, Iowa State University, Ames, IA  50011}
\author{S. Rosenkranz}
\affiliation{Materials Science Division, Argonne National Laboratory, Argonne, IL 60439}
\author{M. R. Norman}
\affiliation{Materials Science Division, Argonne National Laboratory, Argonne, IL 60439}
\author{M. Randeria}
\affiliation{Department of Physics, The Ohio State University, Columbus, OH  43210}
\author{Z. Z. Li}
\affiliation{Laboratoire de Physique des Solides, CNRS, Universit\'e Paris-Sud, 91405 Orsay Cedex, France.}
\author{H. Raffy}
\affiliation{LLaboratoire de Physique des Solides, CNRS, Universit\'e Paris-Sud, 91405 Orsay Cedex, France.}
\author{J. C. Campuzano}
\affiliation{Department of Physics, University of Illinois at Chicago, Chicago, IL 60607}

\begin{abstract}
The loss of single-particle coherence going from the superconducting state to the normal state in underdoped cuprates is a dramatic effect that has yet to be understood.  Here, we address this issue by performing angle resolved photoemission spectroscopy (ARPES) measurements in the presence of a transport current.  We find that the loss of coherence is associated with the development of an onset in the resistance, in that well before the midpoint of the transition is reached, the sharp peaks in the ARPES spectra are completely suppressed.  Since the resistance onset is a signature of phase fluctuations, this implies that the loss of single-particle coherence is connected with the loss of long-range phase coherence.
\end{abstract}

\date{\today}
\maketitle

\section{INTRODUCTION}

In the classic theory of superconductivity of Bardeen, Cooper and Schrieffer \cite{bcs}, an underlying assumption
is the presence of sharp quasiparticles in the normal state.  In underdoped cuprates, this condition is violated in that
the pseudogap phase is associated with broad, incoherent electronic excitations \cite{Ding,Loeser,Norm98}.  
If spectral broadening arises from electron-electron scattering, then it might be tempting to argue that
the closing of the superconducting gap leads to loss of single-particle coherence, 
since scattering processes that were gapped out below T$_c$ could become important above T$_c$.  
But precisely the opposite is seen in experiments.  
In overdoped cuprates where the superconducting gap indeed closes at $T_c$, 
coherent quasiparticles are seen to persist for temperatures well above T$_c$ \cite{adam03}.  
But in the pseudogap phase, coherence is absent above T$_c$ despite the presence of a large energy gap 
that persists to a much higher temperature, T$^*$ \cite{utpal11}.

A number of years ago, it was noted that the intensity of the quasiparticle peak increases upon 
cooling below T$_c$ in underdoped cuprates~\cite{MohitNK,Fedorov} 
and its spectral weight tracks the superfluid density \cite{feng01,ding01}, but the exact relation between 
a two-particle correlation function (the superfluid density) and a single-particle one (the presence of quasiparticles) 
is far from obvious. 

Here, we advocate a new approach to study this important problem by performing 
ARPES measurements in the presence of a transport current \cite{Altman} that induces a 
resistive state in the sample below T$_c$ \cite{Gray, Kanigel, current1, current2, current3}. 
The idea here is to use current flow to destroy the superconducting state, 
distinct from simply raising the temperature above $T_c$, and then use spectroscopy to probe the 
question of single-particle coherence of the electronic excitations. 

Our main result is that the loss of coherent electronic excitations in underdoped cuprates is associated with onset of
resistance, with the sharp peaks in the ARPES spectra completely suppressed well before the midpoint of the resistive transition.  
We argue that the onset of resistance occurs due to motion of vortices in a state with local superconducting order, and 
thus the loss of sharp quasiparticles is tied to the loss of long-range phase coherence in the superconductor.

Before getting into details of our analysis and its implications, we should note that the methodology introduced here
-- ARPES in the presence of current flow -- has the promise of opening up new opportunities 
for probing quantum materials. The investigation of  non-equilibrium states of 
quantum matter is still in its infancy, with pump-probe spectroscopy (optics and ARPES) being
the most commonly used technique for solid state systems.
The new methodology we develop here can be used to probe the single-particle spectroscopy
of non-equilibrium steady states in the presence of current flow. 
This could lead to new insights into many different problems, for instance:
(i) superconducting materials, especially since many
applications necessarily involve current flow; (ii) charge density wave (CDW) materials,
where current flow leads to a depinning of the CDW; (iii) correlated materials with
complex phase diagrams, where current flow could alter the relative stability of competing phases.
 
 The rest of the paper is organized as follows. Given that we introduce a completely new methodology,
a significant fraction of the paper is devoted to a careful discussion of experimental issues. 
 In Section II, we describe the samples and the device geometry. In Section III A we estimate various effects
 related to current flow in the sample, including the effect of the resulting fields on ARPES and the important question of Joule heating.
(with details relegated to Appendices A through E). We note that the analysis of Joule heating 
is central to our work, since only then can we prove that we observe the nontrivial effects of
current flow below the sample $T_c$, rather than the known effects of heating the sample above $T_c$.
We should also emphasize that our ARPES measurements were taken when the resistance of the samples with current flow was only one third of the ``zero-current" resistance above T$_c$, thus Joule heating was small enough not to raise the sample temperature above T$_c$. In Section III B, we present the ARPES data in the present of current flow and contrast the behavior
observed in underdoped and overdoped cuprates. Finally we conclude with some remarks on the broader implication  
of our results for cuprates.

\section{EXPERIMENTAL DETAILS}

We utilize thin ($\sim$500 \AA) films of Bi$_2$Sr$_2$CaCu$_2$O$_{8+\delta}$ (Bi2212) prepared by RF sputtering on SrTiO$_3$ (STO) substrates. These samples possess ARPES and transport characteristics very similar to those of single crystals, but their small cross sections allow us to obtain high current densities of $\sim$10$^6$ A/cm$^2$ using modest values of the current ($\le$200mA). The films also display small signals from the  structural superlattice distortion ($<$3\% intensity of the main band), thus simplifying the interpretation of ARPES data near the Brillouin zone boundary. The thin film samples were patterned into the shape of two large rectangular pads (3 mm by 2 mm) connected by a narrow bridge of width $\sim$250 $\mu$m and length $\sim$1 mm as shown in Fig.~1a,b. Two electrical contacts were made by evaporating  gold onto those two pads on the top of the sample and then attaching a single copper wire to each with silver paste. Current was injected through such made contacts. The residual resistance of the sample-gold junction was measured off-situ in a four point contact setup to be less than 20 m$\Omega$.  The current path is returned parallel to the sample in order to reduce  the magnetic field and provide a ground plane to induce a uniform current through the sample as shown in Fig.~1a. 
The electrical insulation between the sample and the current return electrode was provided by the STO substrate.  The substrate is non-conducting with a resistance larger then 40 M$\Omega$ (limit of our ohm meter) and thus does not contribute to the electrical transport.  The two wires carrying the sample current were thermally anchored to the cold finger, then attached to two pairs (one pair for supplying the current to the sample and one pair for measuring the voltage drop) of thinner, long wires that were wrapped many times around both stages of the cold finger and connected to the electrical feedthrough. Such a configuration, illustrated in the inset of Fig.~1c, minimizes the heat transport from the electrical feedthrough that is at room temperature through the wires to the sample, while minimizing the heat dissipation in the wires attached to the gold pads and maintaining the ability to measure the voltage relatively close to the sample. The power supply operating in constant voltage mode was connected to the current leads and a digital voltmeter was connected to the voltage leads. The IV curves were measured with the UV beam switched off and we did not observe any changes in the current nor voltage values when the UV beam was switched on for the ARPES measurements. A small aluminum pin of similar shape was glued to the top of the bridge and used to cleave its surface in-situ. The thickness of the bridge and thus its resistance will vary from sample to sample due to cleaving and is roughly of the order of 500 \AA.  

ARPES measurements were carried out using our SES50 movable electron energy analyzer and 4m normal incidence monochromator on the U1 undulator beamline at the Synchrotron Radiation Center in Wisconsin. A moveable analyzer allows the acquisition of data where the energy gap is maximal (antinodal regions of the zone), and a chemical potential reference where the gap is zero (zone diagonals), without moving the sample with respect to the photon beam.  The sample is mounted in the geometry where the Cu-O bond direction is parallel to the polarization plane of the photons, with a photon energy of 22 eV employed.

\section{RESULTS AND DISCUSSION}

\begin{figure}
\includegraphics[width=\hsize]{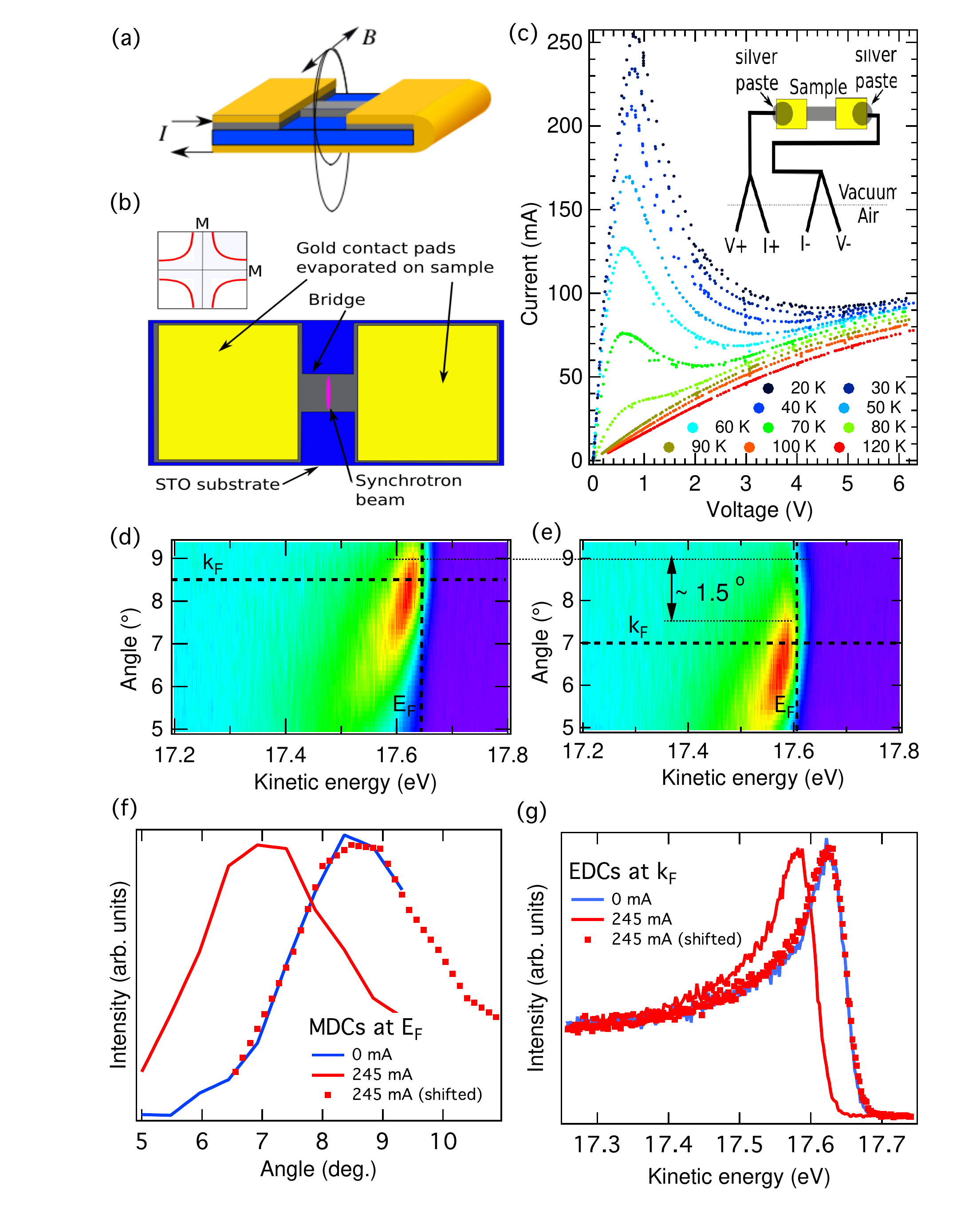}
\caption{(Color online) 
(a) Schematic diagram of the measurement geometry. Gray area is the thin film Bi2212 sample, blue color marks the STO substrate, and gold color signifies the metallic contacts. 
(b) Schematic drawing (not to scale) of the sample, bridge, gold pads and synchrotron beam.  The inset is the 2D Brillouin zone, with M $\equiv$ ($\pi$,0).
(c) $IV$ characteristics at various temperatures for OP Bi2212 sample. The inset shows the electrical connection to the sample inside the vacuum.
(d, e) ARPES intensity along the zone diagonal without (d) and with (e) a current flowing through a copper plate placed underneath the substrate.  A small shift of $\sim$1.5$^{\circ}$ is primarily due to the magnetic field generated by the current (0.845 V, 245 mA). Note that in this test configuration, the current is not flowing through the sample. The shift in the energy is due to the sample being in electrical contact with one of the current leads. E$_F$ and k$_F$ for each case are marked by dashed lines.
(f) MDCs at E$_F$ with and without current, as in (d) and (e). A copy of the MDC with current (dotted curve) is shifted in momentum for a line shape comparison.
(g) EDCs at k$_F$ with and without current, as in (d) and (e). A copy of the EDC with current (dotted curve) is shifted in energy for a line shape comparison. 
}
\label{fig1}
\end{figure}

\subsection{ARPES spectroscopy in presence of current}

When current is flowing in the sample, electric and magnetic fields exist in the vacuum, which deflect the outgoing photoelectrons. We test the effects of these fields empirically by passing a substantial current through a copper plate insulated from the sample and mounted just underneath. The effects of the current are illustrated in Figs.~1d and 1e, where a $\sim$1.5$^{\circ}$ deflection is observed with an applied current. The MDCs at E$_F$ and EDCs at k$_F$ for data with and without the current flow are shown in Figs.~1f and 1g. E$_F$ was determined by integrating EDCs along the momentum cuts and fitting resulting curve with a Fermi function. We do not observe significant distortions of the spectra (other than shifts) that can be created by the highly nonuniform fields. From the magnitude of the shift, sample to analyzer distance, and electron kinetic energy, we place an upper limit on the magnetic field of 1 Gauss close to the sample surface. The EDCs are slightly shifted in energy as the sample is grounded via one of the current leads. The potential drop caused by the flow of current requires that the Fermi level -- the zero of binding energy -- be known at the point of measurement. This is achieved by measuring a reference spectrum at the d-wave node (Fermi crossing along the Brillouin zone diagonal), where the spectral gap is known to vanish at all temperatures, without moving the sample with respect to the photon beam. The leading edge of this gapless spectrum determines the zero of binding energy. 
If the potential drop were to occur at discrete weak links \cite{Gray} in the superconductor, then we would expect to see multiple images of the spectrum displaced in voltage. We do not see any evidence for this in our data. On the other hand, if the potential drop occurs more or less uniformly across the sample, then we would see an inhomogeneous broadening of the ARPES spectrum, which is essentially equivalent to a degrading of the energy resolution. To minimize this broadening, we focus the photon beam to a fine spot $\sim$20 $\mu$m in size along the current direction.  (Detailed considerations of effects of current, voltage and magnetic field in ARPES experiment are discussed in Appendix A).

One might also wonder if the large current density in the sample disturbs the electronic states. A simple Drude model estimate indicates that the change in momentum  of the electrons due to the applied electric field is of order 10$^{-4}$ of the Fermi momentum in the normal state, too small to be  measured (as discussed in Appendix B). Another aspect of these experiments is Joule heating, once the sample enters the resistive regime \cite{skocpol,Zavaritsky}. Heating effects are discussed in detail in Appendix C and D.

\begin{figure}
\includegraphics[width=\hsize]{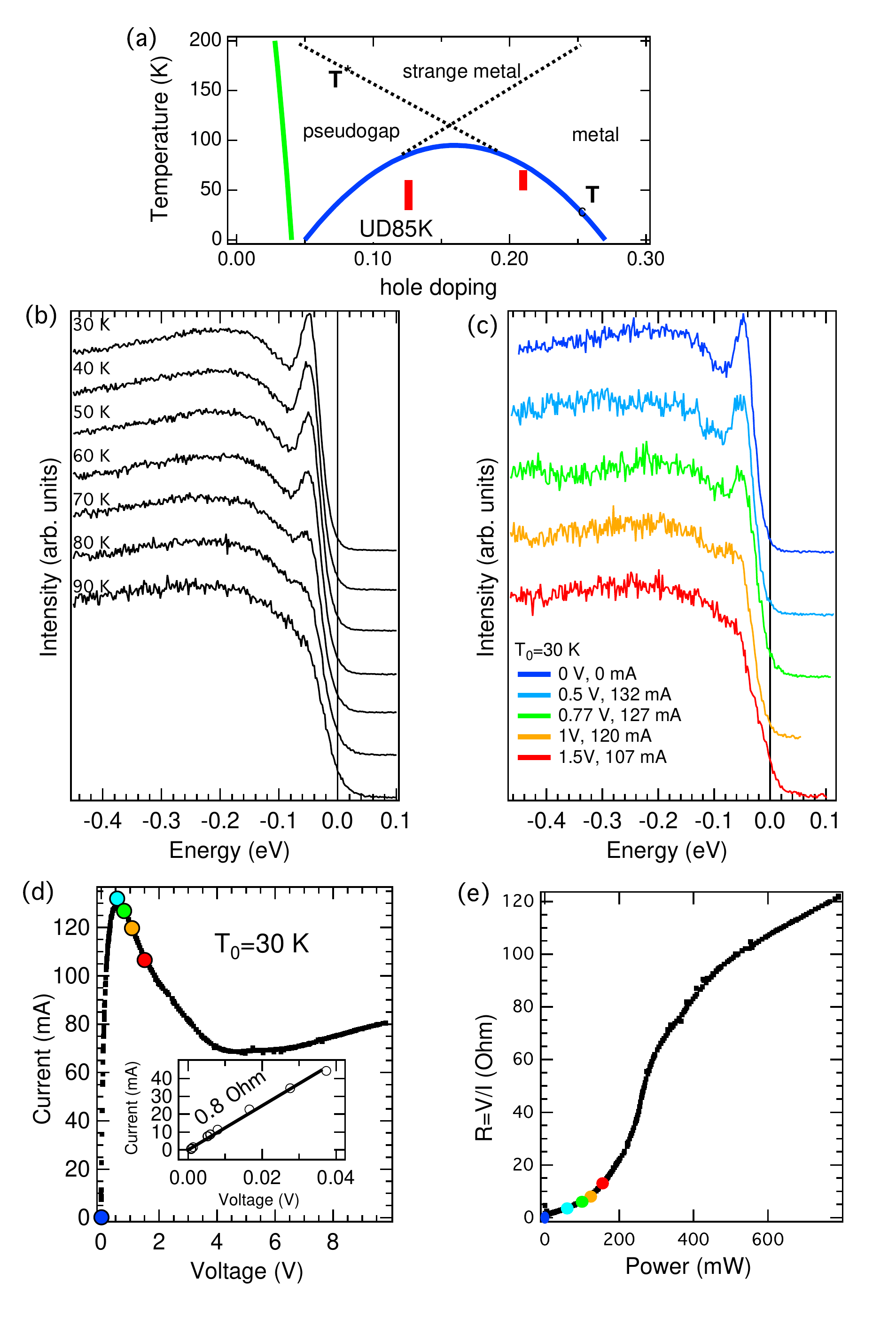}
\caption{(Color online) Spectra at $(\pi,0)$ and $IV$ characteristics for an underdoped T$_c$=85K sample. 
(a) Schematic phase diagram (after Ref.~\cite{utpal11}) showing the doping and temperatures where current data were acquired (indicated by red bars).
(b) $T$ dependence of the ARPES spectrum without current flow. 
(c) Low base temperature (30K) spectrum with current passing through the sample for various voltages.
(d) $IV$ curve for this sample. Colored circles indicate points where the ARPES data were acquired in (c). The inset shows a magnification of the low voltage region. Since the sample is superconducting, the ratio of the voltage to current is equal to the resistance of the wiring and contacts, which for this sample is 0.8 Ohm, much smaller than the normal state resistance.
(e) V/I versus the dissipated power. The colored circles mark points at which ARPES spectra were acquired in (c).}
\label{fig2}
\end{figure}

\subsection{Contrasting behavior of underdoped and overdoped cuprates}

A typical set of current-voltage ($IV$) curves for selected temperatures for an optimal doped (T$_c$=90K) sample is shown in Fig.~1c. 
We use a constant voltage mode to prevent thermal runaway during the transition to the normal state. 
The curves are labeled by the temperature $T_0$ of the cold finger. 
Above T$_c$, at higher voltages the $IV$ curve deviates from a straight line (ohmic) behavior due to heating. This deviation allows us to estimate the sample temperature when the voltage is applied (see Appendix C and D).  Below T$_c$, upon application of voltage, the current increases sharply and its slope is limited by the resistance of the in-vacuum wiring and contacts (we use a two point contact method due to technical limitations). Coincidentally, this helps to limit the rapid onset of the current with voltage in the superconducting state and results in smooth $IV$ curves. 
When the current reaches a critical value, it peaks and then decreases with increasing voltage \cite{Vodolazov}. A crucial question is to what extent this negative differential resistance regime in the $IV$ characteristics arises simply due to Joule heating \cite{Zavaritsky}. 
This has direct implications on whether the effects seen in the ARPES data in the 
presence of current flow are entirely due to heating effects, or 
if they are related to an interesting low temperature resistive state generated by phase slips.

From the analysis presented in detail in Appendix C and D, we conclude that there indeed is an increase in the sample temperature above that of the cold finger, nevertheless Joule heating alone cannot account for all of our observations. Specifically, we conclude that the increase in temperature in the presence of current flow still leaves the sample below T$_c$. This conclusion is based on analyzing the data using two separate methodologies and the simple fact that in this state, the resistance of the samples is significantly below the value measured in the normal state for low currents. In Appendix C, we use a ``pure heating model'', which makes the worst-case assumption that the $IV$ is entirely dominated by heating. 
We show that this model is able to describe many aspects of the data, but not all. We argue that its shortcomings 
imply that this model overestimates the increase in temperature in the regime of interest. 
Next, in Appendix D, we directly estimate the rise in temperature using the measured $IV$ characteristics of the Bi2212 sample, 
together with the $IV$ of a thin layer of gold whose resistivity is similar to the normal state of Bi2212. 
We show that the estimated temperature remains well below T$_c$ in the regime of interest for our ARPES data.

\begin{figure}
\includegraphics[width=\hsize]{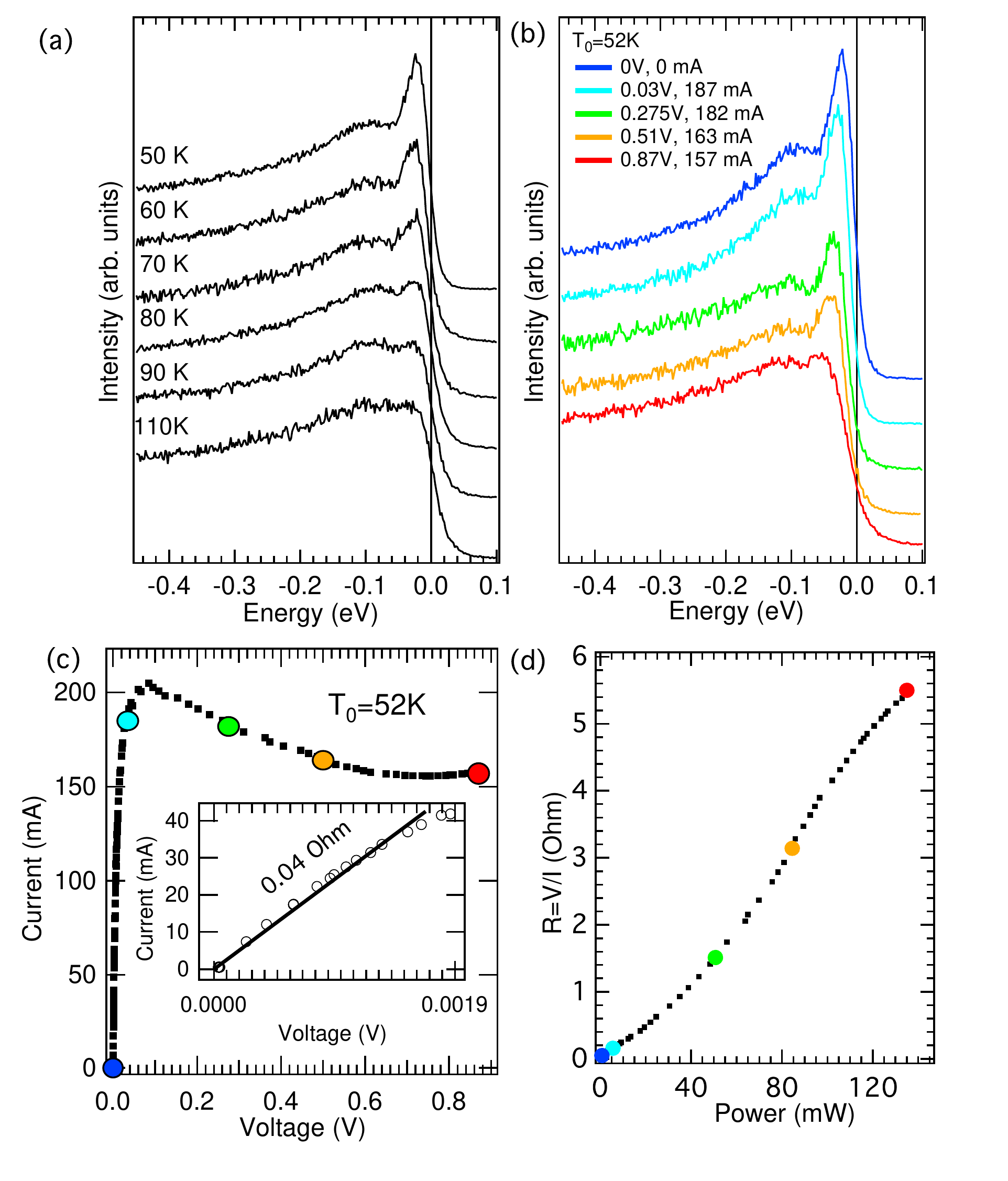}
\caption{(Color online) Spectra at $(\pi,0)$ and $IV$ characteristics for an overdoped T$_c$=75K sample. (a) $T$ dependence of the ARPES spectrum without current flow. 
(b) Low base temperature (52K) spectrum with current passing through the sample for various voltages.
(c) $IV$ curve for this sample. Colored circles indicate points where the ARPES data were acquired in (b). The inset shows a magnification of the low voltage region. Since the sample is superconducting, the ratio of the voltage to current is equal to the resistance of the wiring and contacts, which for this sample is 0.04 Ohm, much smaller than the normal state resistance.
(d) V/I versus the dissipated power. The colored circles mark points at which ARPES spectra were acquired in (b).}
\label{fig3}
\end{figure}

In Fig.~2a, we plot a schematic phase diagram of the cuprates and mark the locations where ARPES data was measured. In Fig.~2b, we show the ARPES spectrum of an underdoped (T$_c$=85K) sample at the $(\pi,0)$ point of the Brillouin zone as a function of the temperature. Upon increasing the temperature, the quasiparticle peak decreases in intensity and vanishes close to T$_c$ \cite{MohitNK,Fedorov,feng01}, while the pseudogap persists well above T$_c$ \cite{ding01,Loeser}.  At a low cold finger temperature, we drive the current through the sample by applying a voltage and measure the spectrum at the $(\pi,0)$ point as shown in Fig.~2c. We use the spectrum at $(\pi,0)$ rather than k$_F$ because it is easier to establish the presence of the coherent peak. The chemical potential is determined at each voltage by measuring the nodal spectrum, where the superconducting gap is zero. The current/voltage values for each ARPES measurement are color coded on the $IV$ curve in Fig.~2d. The top spectrum in Fig.~2c was measured without the current and is used as a reference. We start close to the peak of the $IV$ curve (Fig.~2d), which corresponds to the critical current  for this cold finger temperature.  The corresponding ARPES spectrum looks very similar to the reference, with a pronounced quasiparticle peak and superconducting gap. Since this spectrum was measured at the highest value of current, its similarity to the reference spectrum demonstrates that the magnetic field does not significantly affect the ARPES line shape. The presence of the coherent peak and superconducting gap in the spectrum also clearly demonstrates that the peak of the $IV$ curve corresponds to the vortex depinning \cite {MFMdepinning} critical current  rather than the depairing current. The ratio of the current flow velocity to the depairing velocity is very small, thus the flow velocity (estimated to be at most $\sim$30 m/s) is two orders of magnitude smaller than the depairing velocity $\Delta/\hbar k_F \sim$9000 m/s, where $\Delta$ is the energy gap and $k_F$ the Fermi momentum  (see Appendix B). Surprisingly, as we increase the voltage, the quasiparticle peak decreases rapidly in intensity and it vanishes at a point where the current is significantly higher than the value observed above T$_c$. Even taking into account the heating (see Appendix D), we estimate the sample is still at $\sim$60K, well below T$_c$, at which temperature the coherent peak is still present in the absence of the current (Fig.~2b).
Thus, the loss of coherence observed in Fig.~2c corresponds to the onset of the resistance, rather than simply entering the normal state
(i.e. due to heating alone), as is made clear in Fig.~2e, which shows the variation of the resistance (V/I) as a function of the power (IV).  
Clearly, the flow of current leads to extra dissipation, which in turn destroys single-particle coherence.
Presumably, this extra dissipation is due to the development of phase slips and vortices \cite{skocpol,beasley,martin} rather than a loss of the pairing amplitude, as a well defined energy gap is still present at the highest voltage.

We contrast this behavior with one of an overdoped (T$_c$=75K) sample shown in Fig.~3. In Fig.~3a we plot the temperature dependence of the spectrum at ($\pi$,0) measured with no current passing through the sample. The ARPES spectra shown in Fig.~3b are measured for several current/voltage values indicated on the $IV$ curve in Fig.~3c. With increasing voltage, the quasiparticle peak decreases in intensity, but it remains visible even at the highest value of the voltage, where we have reached the normal state by a combination of the current flow and 
sample heating (the resistance of the sample at 0.87 V, Fig.~3d, being comparable to the dc resistance just above T$_c$). 

\section{CONCLUSIONS}

In summary, by measuring ARPES in the presence of a transport current, we have found that for underdoped samples, the loss of the quasiparticle peak occurs before reaching the normal state. Since the additional dissipation below T$_c$ due to current flow is thought to be due to phase slips and vortices, this indicates that superconducting phase fluctuations destroy the single-particle coherence, a very non BCS-like behavior.  

Our findings are of relevance to a microscopic understanding of high T$_c$ superconductivity in the cuprates. Different ways of destroying superconductivity in the cuprates seem to lead to different ``normal" states. For instance, raising the temperature above T$_c$ in zero magnetic field leads to a non-Fermi liquid state, with a pseudogap near the antinode with Fermi arcs near the nodes, whose origin is still being debated \cite{Keimer}. On the other hand, turning on a high magnetic field at low temperature leads to a Fermi liquid state 
with a Fermi surface reconstructed by broken translation symmetry \cite{dhva}. 

Here we have proposed a third route to destroying superconductivity by passing a current through the sample, and addressed the question of single-particle coherence. In the future, it will be interesting to fully elucidate the nature of the resistive state arising from current flow.
The experimental approach presented above could be further developed by pulsing the current instead of working in a constant voltage mode, thus minimizing the heating \cite{anagawa,Lang}.
We hope to report on such challenging experiments in the future.

\section{ Acknowledgments} 
Work at Ames and Argonne was supported by the Materials Sciences and Engineering Division, Basic Energy Sciences, Office of Science, U.S. DOE. 
M.R. was supported by DOE-BES Grant No.~DE-SC0005035.
This work was  based in part on data acquired at the Synchrotron Radiation Center, University of Wisconsin-Madison, which was supported by the National Science Foundation under Award No. DMR-0537588.

\appendix

\section{CHANGES OF THE KINETIC ENERGY AND MOMENTUM OF THE PHOTOELECTRONS DUE TO THE CURRENT AND VOLTAGE} 

\begin{figure}
\includegraphics[width=\hsize]{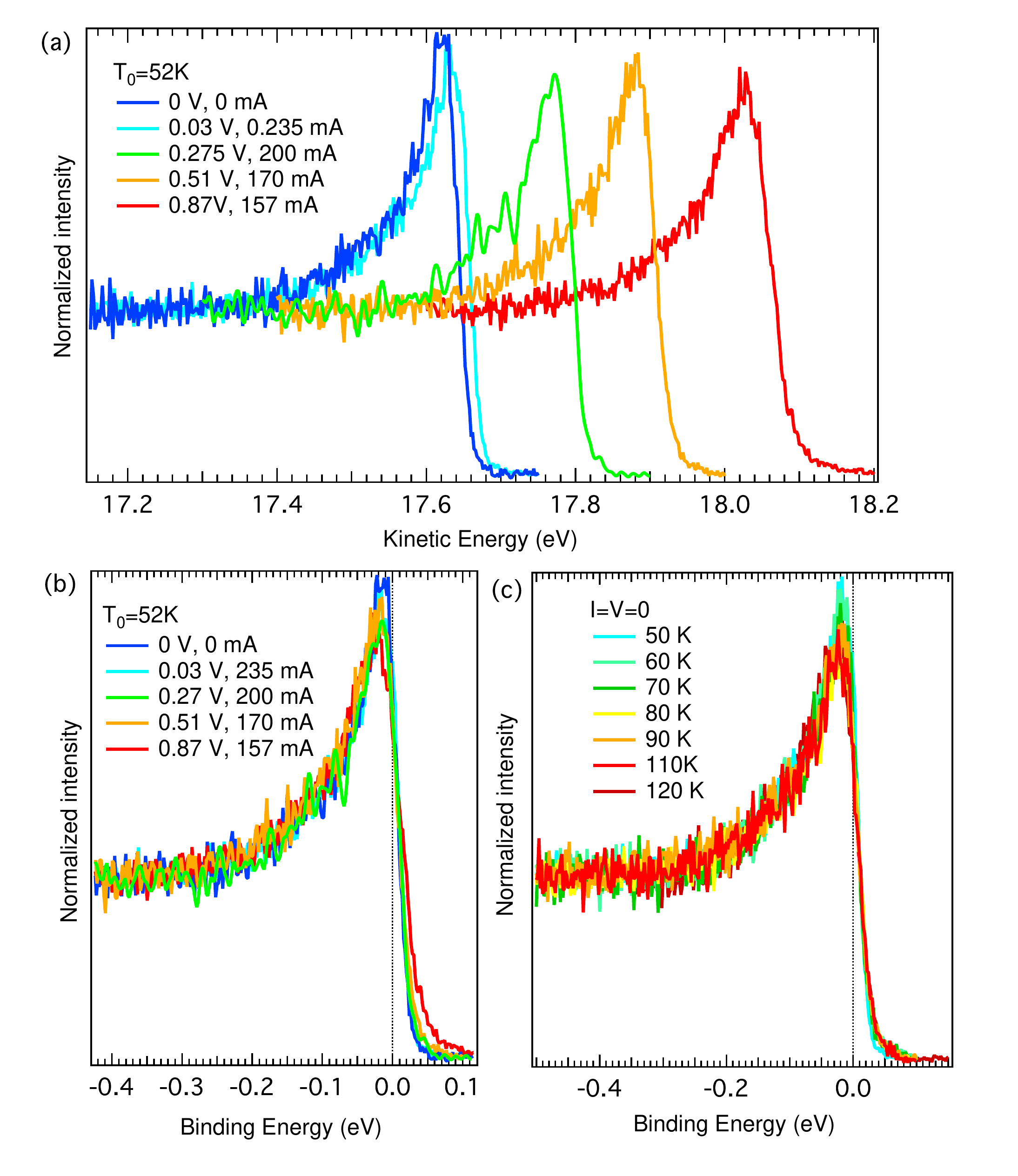}
\caption{(Color online) EDCs at k$_F$ along the nodal direction for several voltages applied across the same sample as in Fig.~3, measured at a cold finger temperature T$_0$=52K. (a) EDCs plotted as a function of the kinetic energy without any offsets. (b) EDCs plotted vs the binding energy. E$_F$ for each measurement was obtained by integrating the EDCs along the momentum cut and fitting the result with a Fermi function.
(c) Zero current EDCs as a function of temperature.  Note the contrast in the leading edge relative to that shown in (b).}
\label{fig4}
\end{figure}

The presence of electric and magnetic fields between the sample and the electron analyzer affects the trajectories of the photoelectrons. In Fig.~1 we demonstrated that the magnetic field changes the angle of the photoelectrons without significantly affecting the spectral line shape. This can be also seen in Figs.~2c and 3b, where the line shape at the peak current value is very similar to the one measured with no current. 

 To minimize the effects of the magnetic field, the return current is routed under the sample. For a single current element, the magnetic field decreases with distance as 1/r$^2$. Since we use two such elements with opposing currents, the resulting magnetic field decreases even faster ($\sim$1/r$^3$). For example, if a 6 Gauss field exists 1 mm from the sample, then 5 mm away the field is $\sim$100 times smaller (i.e. $\sim$0.04 Gauss) and a cyclotron radius $\sim$100 times larger (i.e. 2 m), 10 mm away this field is $\sim$1000 times smaller and a cyclotron radius 1000 times larger (i.e. 20 m). We therefore do not expect significant effects of the magnetic field on the spectra for modest values of the current.

The presence of a voltage drop across the sample has a more significant effect on the spectra. The voltage at a given point of the sample shifts the local chemical potential, which in turn affects the kinetic energy of the emitted photoelectrons. We demonstrate this in Fig.~4a, where we plot EDCs at k$_F$ vs the actual kinetic energy. As the voltage is applied across the sample, the kinetic energy of the photoelectrons increases and the EDCs are shifted to the right. The amount of the shift is equal to the voltage that exists at the position on the sample from which the photoelectrons are extracted. The value of the chemical potential for each voltage value can be obtained from such data measured at the node (where the superconducting gap is zero) by integrating the EDCs along the momentum cut and fitting the result with a Fermi function. The EDCs plotted vs binding energy after correcting for the changes of E$_F$ are shown in Fig.~4b. The second effect of the voltage applied to the sample is a slight broadening of the spectra. This is because the photon beam has a finite size along the direction of the current flow and photoelectrons emitted within the spot size will have slightly different kinetic energies. This effect can be seen in Fig.~4b, where at higher voltage values, there is slight broadening of the leading edge which is not evident in the zero current data, even at elevated temperatures (Fig.~4c). To estimate this effect, let us assume that a potential drop of 1 V occurs uniformly across a 1 mm sample. Then the variation in the chemical potential across the 20 micron beam width is $\delta\mu$ = 20 meV which is a significant effect. This leads to spectral broadening, i.e. $I(k,\omega) = \int d\omega^\prime I_0(k,\omega-\omega^\prime)$ where the integration is over the window  $-\delta\mu/2 < \omega^\prime < +\delta\mu/2$ (here, $I$ is the measured intensity and $I_0$ the intensity without voltage broadening). In a sense, this is similar to the energy resolution convolution, and can be thought of as degrading the energy resolution.
This broadening is roughly proportional to the ratio of the beam size along the direction of the current to the length of the bridge. This is perhaps the most challenging part of this experiment, as it requires the use of a very small photon beam to minimize the broadening effects, which in turn limits the photon flux and requires extended time for data acquisition. 

\section{ESTIMATES OF CHANGES OF KEY QUANTITIES DUE TO FLOW OF CURRENT AND VALUE OF DEPAIRING CURRENT} 

{\bf Shift in $k$ due to current flow:}

The Drude model implies $\delta k = e \tau E / \hbar$.
For $V = 1$ volt across a 1 mm long sample,
$eE = 10^3$ eV/m.
For the relaxation time, one can estimate
$\hbar/\tau = k_B T$, so that at 100 K one gets
$\tau = 10^{-13}$ s.
Since $\tau$ increases below T$_c$,
an upper bound estimate at T=50K is
$\tau = 10^{-12}$ s.  This yields  $\delta k = 1.5 \times 10^6 m^{-1}$.
Using $k_F \simeq \pi/a \simeq 10^{10} m^{-1}$
results in $\delta k / k_F \simeq 10^{-4}$, which is
entirely negligible.

{\bf Flow velocity:}

Assuming a uniform current distribution, a current of 125 mA 
through a cross-sectional area of 0.25 mm $\times$ 500 $\AA$ gives a
current density $J = 10^{10}$ Amps/m$^2$.
Note, we assume the thinnest possible sample so that, if the actual 
sample thickness is more like 1000 to 1500 $\AA$, then the current 
density $J$ may well be smaller by a factor of 2 to 3. To estimated flow velocity is $J = nev$. The appropriate carrier density $n$ to use is not so obvious, but these
uncertainties will not be of any importance as seen below.
If we use the smallest possible estimate $n \sim x$ (doped holes) then $n \simeq 0.25/[3.83 \times 3.83 \times 7.5] \AA^{-3}$ where we used a value of 7.5 $\AA$ as the average spacing between layers in Bi2212, so  $n = 2.25 \times 10^{27} m^{-3}$. This estimate
of $n$ gives an estimated $v \simeq 30 m/s$.
Note that $v/v_F \sim 10^{-4}$ (where the experimental nodal $v_F$ was used). If, on the other hand, we assume 1 carrier per unit cell (as appropriate to $n \sim 1+x$ holes), then $n$ goes up by a 
factor of 5, and correspondingly $v$ is reduced by a factor of 5.

{\bf Depairing velocity:}

As shown in Tinkham \cite{tinkham}, the depairing velocity is is reached when the Doppler shifted spectrum becomes gapless. In our case $v_c = \Delta/p_F$.  Taking $\Delta = 40$ meV and $p_F = \hbar k_F$ ($k_F$ at the antinode is 0.67 \AA$^{-1}$), we get $v_c \simeq 9000$ m/s.
So in the absolute ``worst" case, our flow velocity is over
two orders of magnitude less than the depairing velocity.  A large difference between the drift velocity for the critical current and the depairing velocity is expected and also seen in classical superconductors \cite{Buzea}. This is reassuring because (a) the observed critical current is
much smaller than the maximum possible theoretical value;
(b) if the pseudogap has anything to do with pairing, the current flows should not be large enough to cause depairing.

\begin{figure}
\includegraphics[width=\hsize]{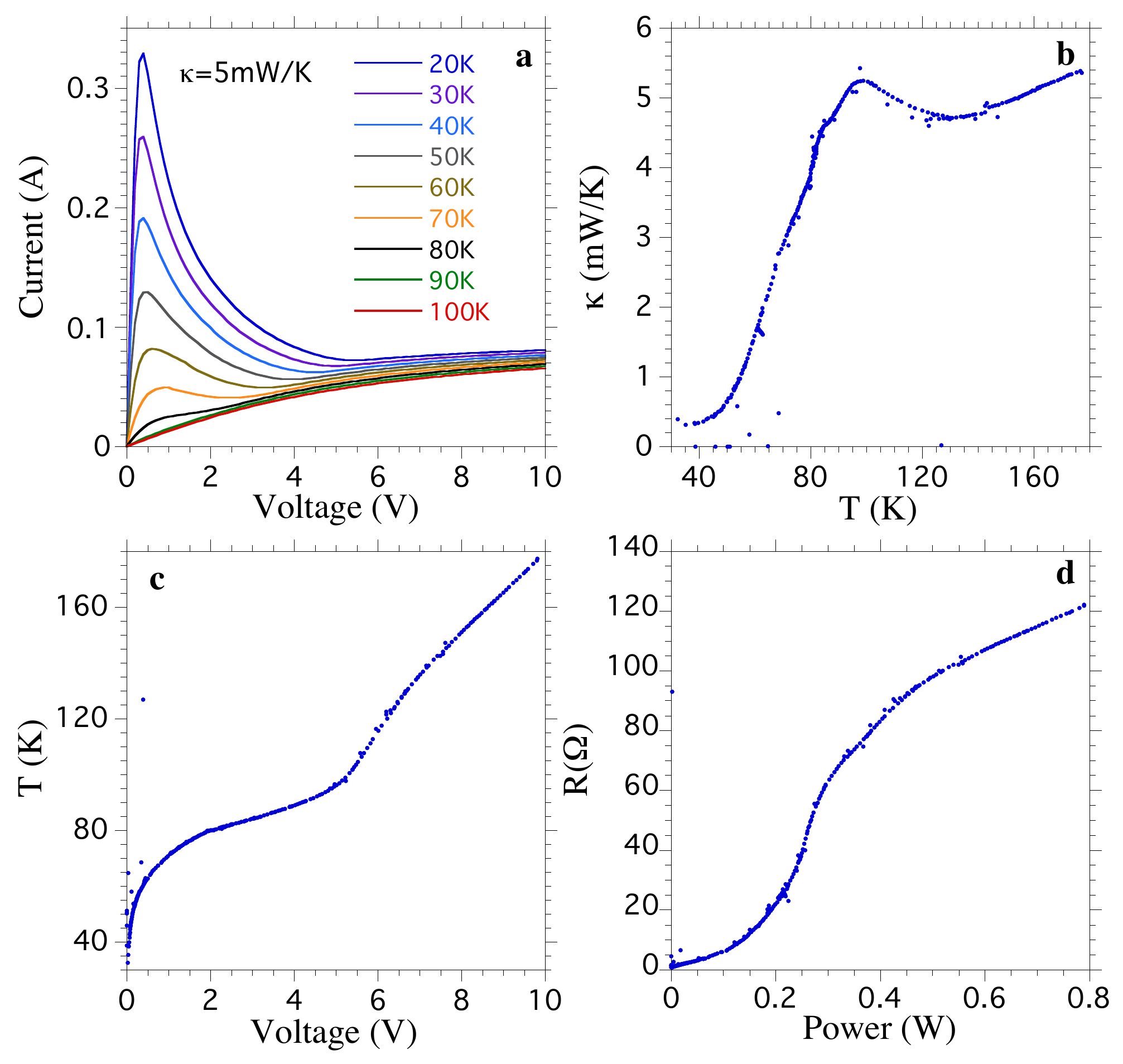}
\caption{(Color online) A pure heating model analysis of the data in Fig.~2.  (a) Simulated $IV$ curves for various base temperatures (assuming a thermal conductance 
$\kappa$ of 5 mW/K) obtained using the dc resistivity curve of the sample in Fig.~2. (b) $\kappa(T)$ obtained by fixing the current $I$ to
that shown in Fig.~2d.  (c) Resulting variation of $T$ with the bias voltage.  (d) Resistivity (V/I) versus power (IV) as in Fig.~2e.}
\label{fig5}
\end{figure}

\section{ESTIMATION OF HEATING EFFECTS: PURE HEATING MODEL}

Applying a voltage across the sample will lead to Joule heating.  Such heating will cause the sample temperature to vary as $T = T_0 + IV/\kappa(T)$, where $T_0$ is the base temperature and $\kappa(T)$ the thermal conductance \cite{Zavaritsky}. For simplicity, let us first assume that the thermal conductance is constant with $T$.  Then, for a given voltage, the current is given by the condition $I = V/R(T_0+IV/\kappa)$, which can be determined by simple root finding.  $R(T)$ is obtained from dc measurements of the resistance of the cleaved sample using a small constant current.  In Fig.~5a, we show simulated current versus voltage curves for various base temperatures using a $\kappa$ of 5 mW/K in order to match the high voltage data in Fig.~2d.  One sees a striking resemblance of these curves to those shown in Figs.~1c and 2d, not only in shape and evolution as a function of base temperature, but also in the voltage location of the current maximum.  The curves are not an exact match with Fig.~2d, though.  In particular, the current maximum is significantly larger in the simulated curves.  In a pure heating model, this would be attributed to the $T$ dependence of $\kappa$.  

To see this, one can fix $I$ from Fig.~2d and extract $\kappa(T)$. This is shown in Fig.~5b, and has some resemblance to the $T$ dependence of the known thermal conductance of Bi2212 \cite{ando2000}, but the inferred $\kappa$ drops by a factor of ten when going from the normal to the superconducting state, unlike bulk Bi2212 which only drops by about a factor of two.  We should remark that the actual thermal conductance should be limited by that of the STO substrate, which has an even milder temperature dependence than Bi2212 \cite{Yu}.  This is verified by doing the same heating analysis for a Au film on STO (see Appendix D) at the same base temperature, which results in a weak temperature dependence for $\kappa$, giving rise to a nearly linear variation of $T$ with power.

From Fig.~5b, one can plot $T$ versus the voltage, as shown in Fig.~5c.  In essence, one is using the sample as a thermometer. But the rapid rise of $T$ at low voltages seems unphysical, and is related to the unphysically large drop in $\kappa$ in Fig.~5b.  This is also evident in Fig.~5d where we plot $R$ versus the power, IV.  The development of resistance with power is more dramatic than the development of the dc resistance with temperature (Fig.~7b).  This implies, as discussed in the main text, that there is extra dissipation below T$_c$ due to the current flow.

\begin{figure}
\includegraphics[width=\hsize]{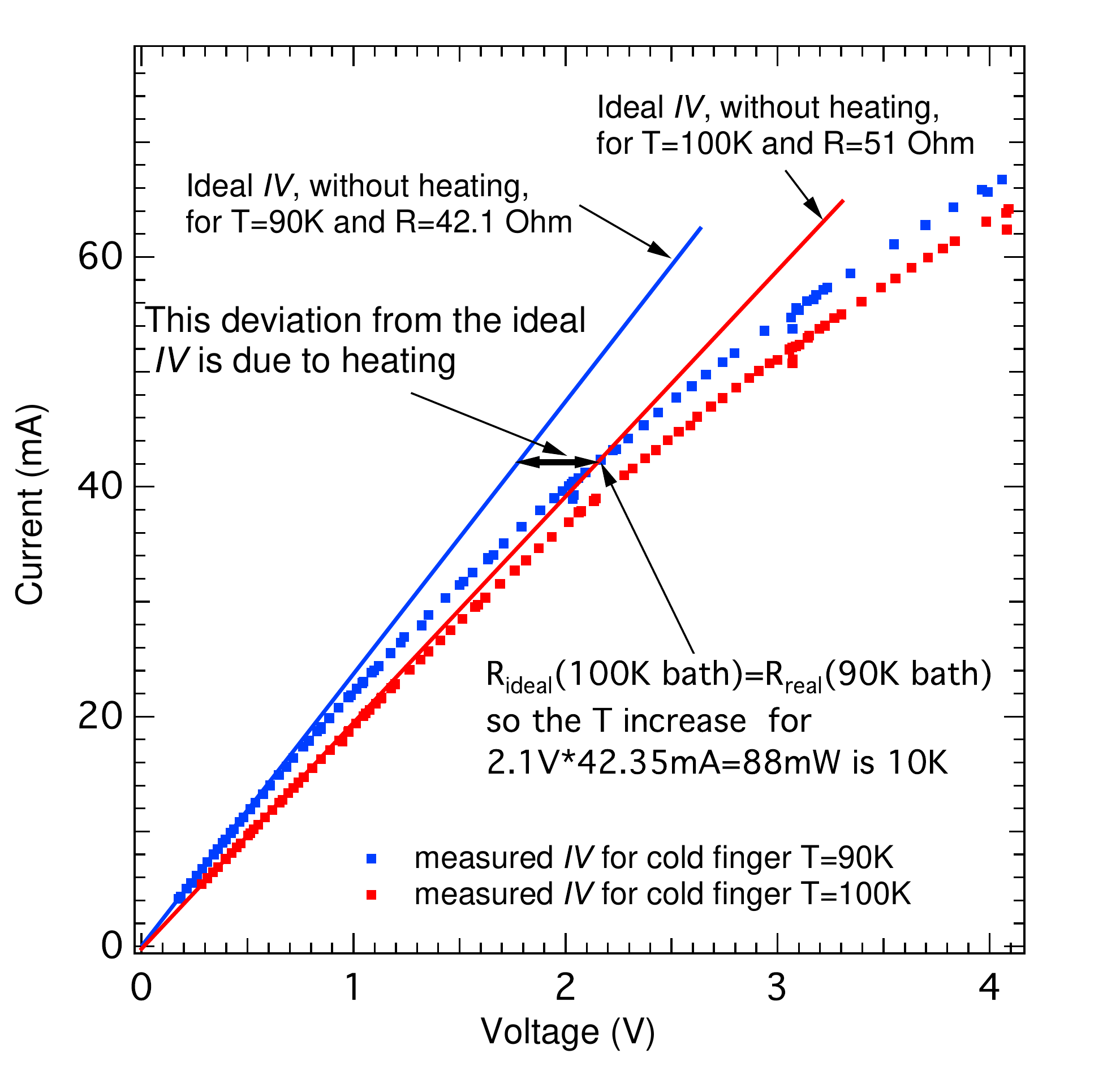}
\caption{(Color online) $IV$ curves above T$_c$ from Fig.~1c. Straight lines are fits to the low voltage part and signify ideal $IV$ curves (with a slope equal to the inverse of the resistance) of the sample in the absence of heating.}
\label{fig6}
\end{figure}

\begin{figure}
\includegraphics[width=\hsize]{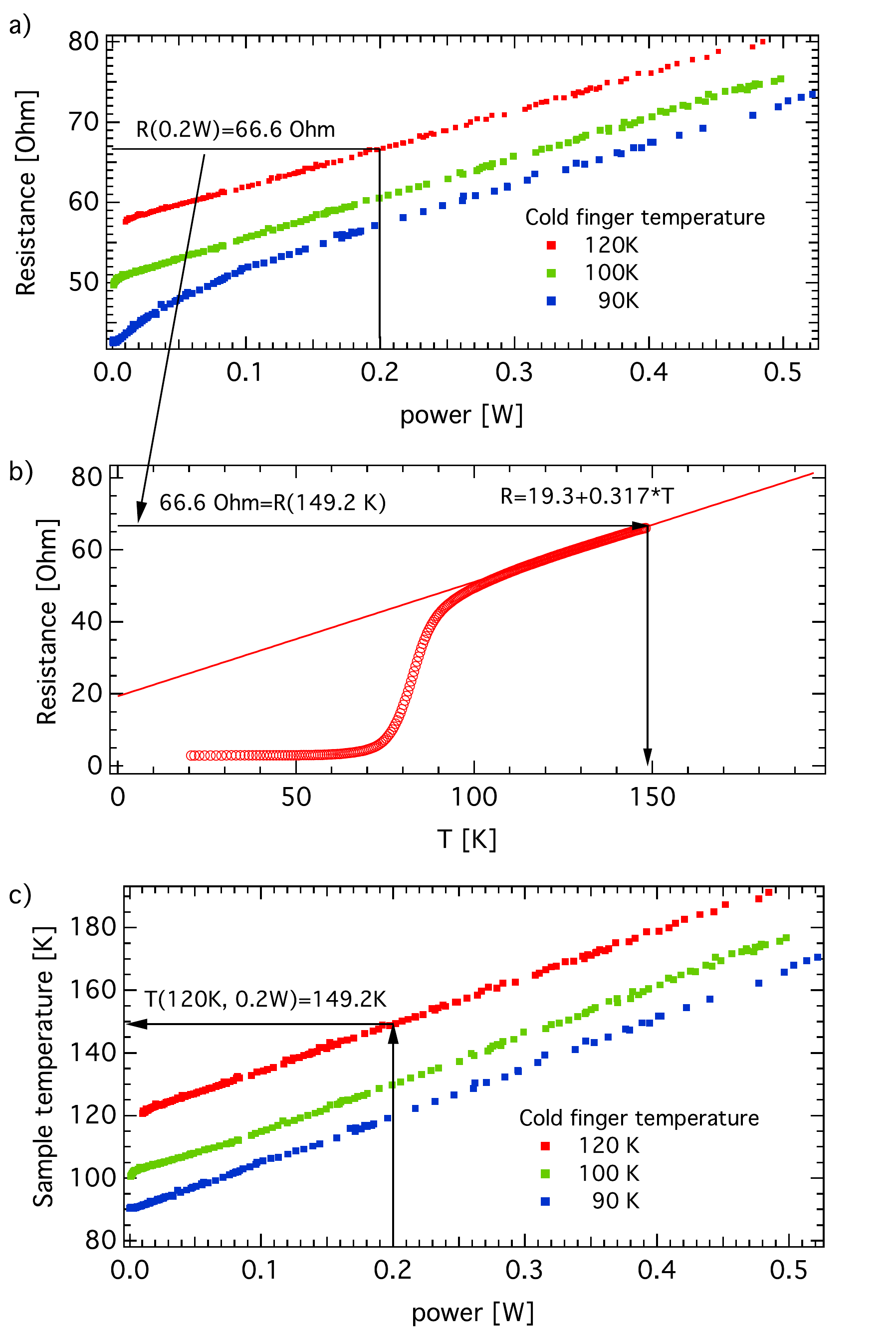}
\caption{(Color online) Estimation of the sample temperature in the presence of voltage using data for OP Bi2212, same as in Fig.~1c. (a) Resistance versus power for three temperatures of the cold finger. (b) dc resistance versus temperature. (c) actual sample temperature versus power calculated by converting the data in (a) using the dc resistance shown in (b).}
\label{fig7}
\end{figure}

\section{EMPIRICAL ESTIMATE OF HEATING} 

The actual sample temperature above T$_c$ under current flow is relatively easy to determine, as discussed in Appendix C. To accomplish this, we use the sample itself as a thermometer, utilizing the temperature dependence of the dc resistance. To illustrate this, we plot the $IV$ curves measured for cold finger temperatures above T$_c$ in Fig.~6. The dotted curves are the actual measured $IV$. The straight lines are fits to the low voltage data and represent ideal $IV$s in the absence of heating, with their slopes being equal to the inverse resistance. The actual $IV$ for T$_0$=90K (blue dots) crosses the ideal T$_0$=100K (red line) at $\sim$2.1V (P=88mW), therefore at this power dissipation the actual sample temperature is T=100K (a heating of 11.4K/100 mW). A more systematic approach is to use the $R$ versus power plot obtained from the $IV$ curves, and then convert the resistance for a given power to temperature using the temperature dependence of the resistance. This is demonstrated in Fig.~7. At 200mW and a cold finger temperature of T$_0$=120K, the resistance of the sample (obtained by dividing the voltage by the current) is 66.6 Ohms (Fig.~7a). This value of the resistance corresponds to a temperature of 149.2K on the $R$ versus $T$ curve shown in Fig.~7b. Therefore, a dissipation of 200mW causes a heating of the sample by 29.2K (14.6K/100 mW), a bit higher than the result at T$_0$=90K. The $R$ vs power curve can be also directly converted to an actual sample temperature versus power. This is done by again utilizing the $R$ versus $T$ curve.  The result is shown in Fig.~7c for various base temperatures, and allows one to directly read off the actual temperature of the sample for any power dissipation within the measured range. 

\begin{figure}
\includegraphics[width=\hsize]{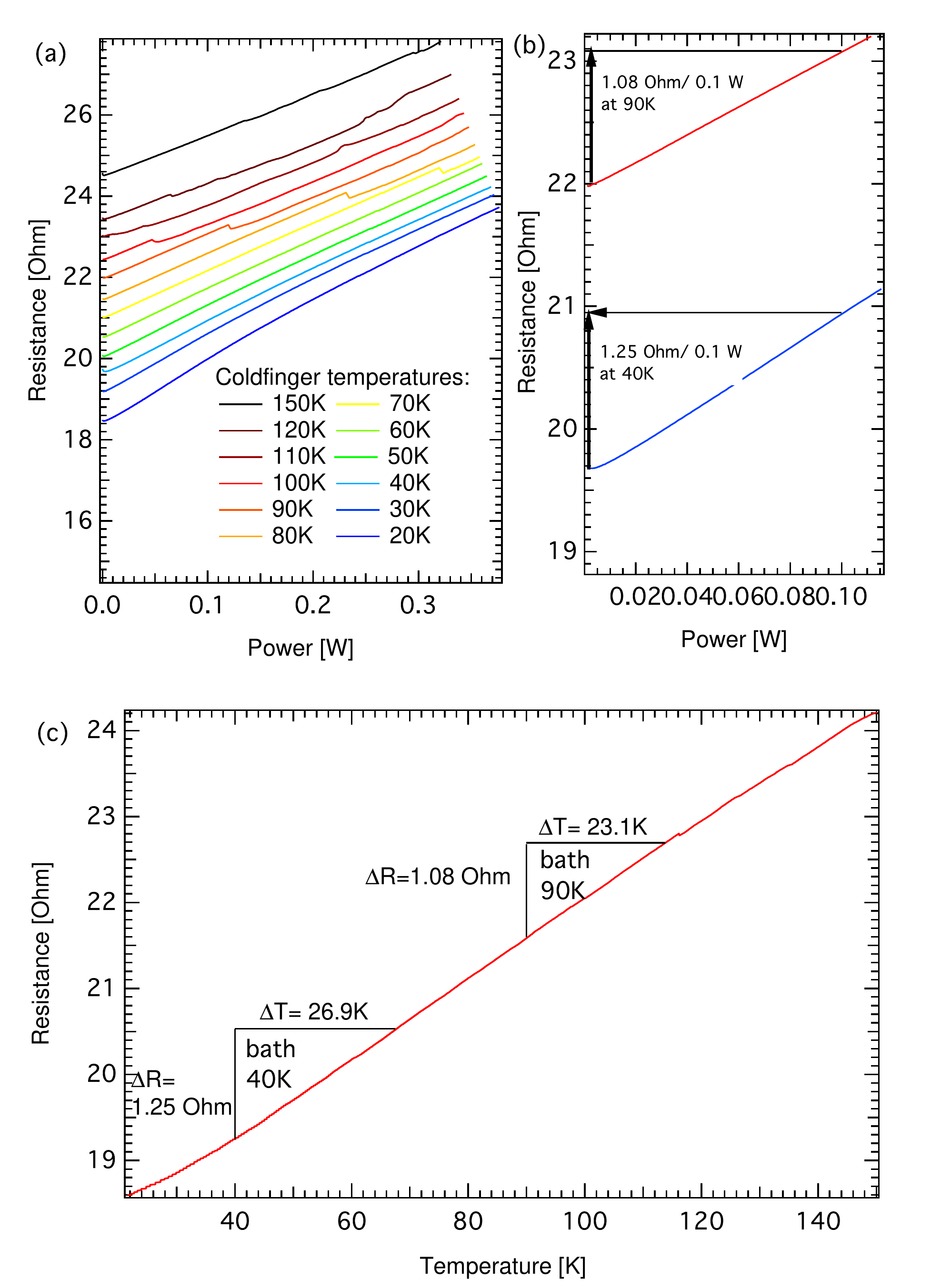}
\caption{(Color online) Heating effects at low and high temperatures in a gold film evaporated on the STO substrate under conditions similar to ones in Fig.~2. (a) $R$ versus power for various cold finger temperatures. (b) $R$ vs power for two temperatures of the cold finger showing the values of the increase of the resistance with power dissipation. (c) $R$ versus $T$ curve used for converting the increase of the resistance to an increase of the temperature.}
\label{fig8}
\end{figure}

The estimation of the sample temperature for cold finger temperatures below T$_c$ is more complicated because of the presence of the superfluid. It is difficult to separate the changes of the resistance due to heating from 
those due to phase slips, 
as discussed in Appendix C. Furthermore, although the thermal conductance of Bi2212 does decrease below T$_c$ \cite{ando2000}, since the thickness of the sample is only $\sim$500 \AA~and therefore 5000 times smaller than that of the STO substrate, we can safely assume that the vast majority of the thermal gradient occurs across the STO substrate, interfaces \cite{SchlomNM} and epoxy with which the substrate is attached to the copper block. The thermal conductivity of STO peaks at $\sim$17~W/m-K at 100~K and decreases to $\sim$10~W/m-K at 40~K \cite{Yu}. The thermal conductance of the epoxy is difficult to locate, but its thickness is smaller than the one of the STO substrate by at least an order of magnitude. Even though the presence of various interfaces, the substrate and the epoxy makes this a complicated heat transfer system, the thermal gradient across it can be easily determined experimentally at low temperatures by utilizing a non-superconducting film. To do this, we carefully stripped the Bi2212 film, cleaned the substrate and evaporated a thin layer of gold so that its resistivity is similar  to the one in the normal state of Bi2212. We can then study the thermal resistance of the substrate at low cold finger temperatures in a similar fashion as described above. In Fig.~8a, we plot the resistance versus power curves for a number of cold finger temperatures. In Fig.~8b, we focus on two values of T$_0$=40K and 90K. At 100 mW dissipation, the increase of the resistance is 1.25 Ohms and 1.08 Ohms, respectively, which can be converted to an increase of the sample temperature using the $R$ versus $T$ curve shown in Fig.~8c. We find that at T$_0$=40K, the sample heating is 27K per 100 mW and for T$_0$=90K, it is 23K per 100 mW. The decrease of the thermal conductance at low temperatures is consistent with data available for STO, where there is a peak in the thermal conductance at $\sim$100K \cite{Yu}.  As mentioned in Appendix C, use of a pure heating analysis leads to a mild increase of the thermal conductance by 18\% at a base temperature of 30K when the power increases from zero to 370 mW.

\begin{figure}
\includegraphics[width=\hsize]{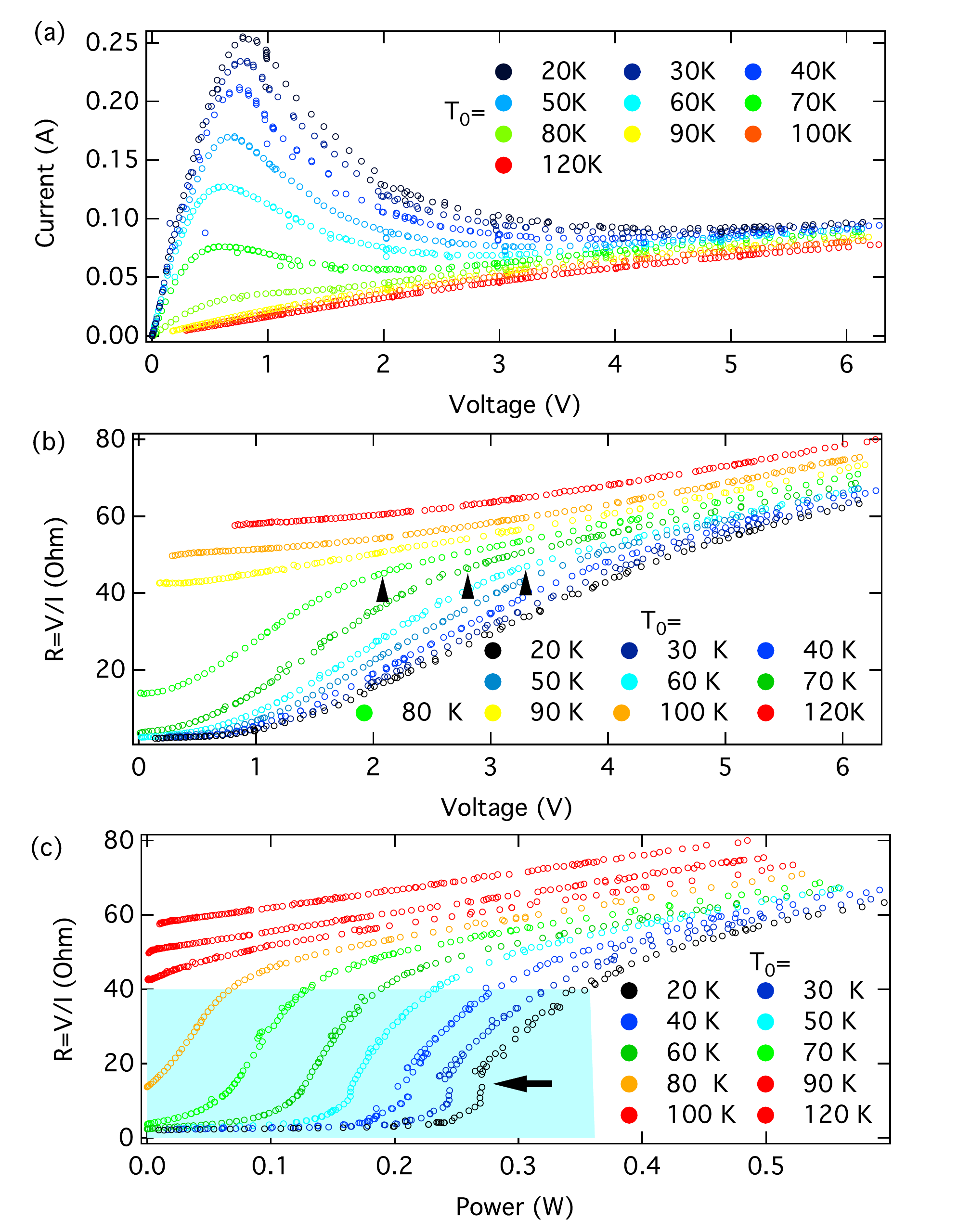}
\caption{(Color online) Transport data for the UD85K sample.
(a) IV curves (same as in Fig.~1c).
(b) V/I ratio vs voltage. Arrows mark a change in slope for the temperatures indicated.
(c)  V/I ratio vs power. The blue shaded rectangle marks the area where the V/I ratio deviates from the normal state linear in power behavior,
indicating the presence of superconductivity.}
\label{fig9}
\end{figure}

In Fig.~9, we compare the IV curves, V/I ratio vs voltage and V/I ratio vs power for the data of Fig.~1. For base temperatures above T$_c$, the V/I vs voltage curves, shown in Fig.~9b, are non-linear due to heating (as demonstrated in Fig.~6). For base temperatures below T$_c$, there is a change of slope (marked by arrows in Fig.~9b) that indicates the presence of superconductivity for lower voltage values. As demonstrated previously in Fig.~7, the V/I ratio for base temperatures above T$_c$ depends linearly on power. Such a plot therefore better reveals the presence of superconductivity at lower voltage values and is shown in Fig.~9c. For base temperatures of 80K and below, there is a deviation from the normal state linear behavior with power even for power values as high as 350 mW.  This drop in the V/I ratio as compared to a linear extrapolation from higher voltages is caused by the presence of superconductivity, and provides evidence that even at such an elevated power, the sample temperature is below T$_c$ within the area marked by the blue shaded rectangle.

\begin{figure}
\includegraphics[width=\hsize]{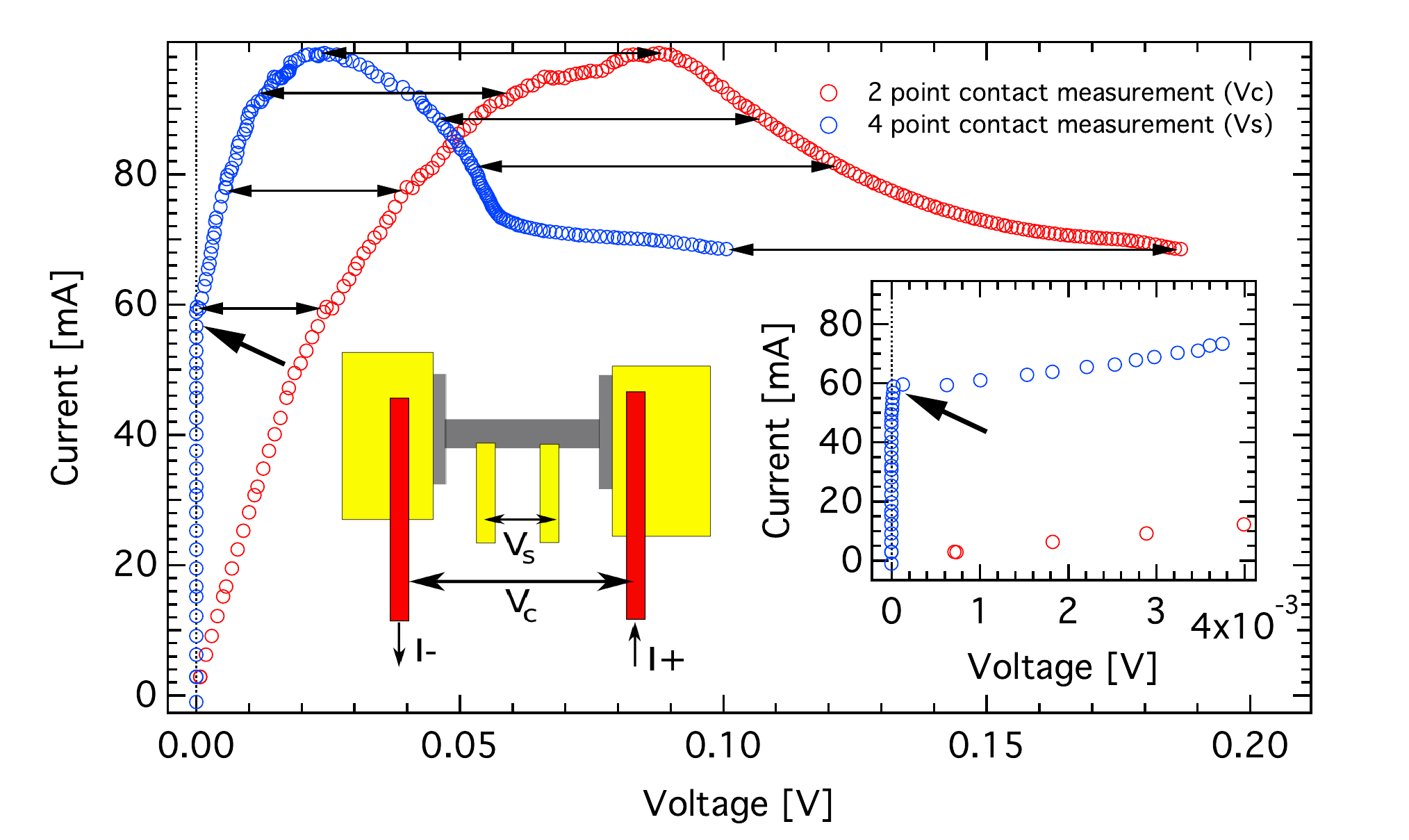}
\caption{(Color online) Comparison of IV curves measured using a 2-point and a 4-point contact method at a cold finger temperature of 70K and an optimally doped sample with T$_c$=90K. The voltage for the red curve (V$_c$) was measured across the outer current contacts and the voltage for the blue curve (V$_s$) was measured across two separate electrodes connected to the mid section of the bridge that do not have direct electrical contact with the current electrodes other than through the sample. The single arrow indicates the value of the current at the onset of the resistive transition.  The black double arrows connect pairs of V$_c$ and V$_s$ points that were measured at the same time. The left inset shows a drawing of the measurement geometry. The sample is shown in gray, gold contacts in yellow and copper leads in red. The silver paste linking the copper leads and gold contacts is not shown for clarity. The right inset shows the expanded low voltage region.}
\label{fig10}
\end{figure}

\section{COMPARISON OF 2-POINT CONTACT AND 4-POINT CONTACT MEASUREMENTS OF THE IV CURVES} 

Due to technical limitations, we utilized 2-point contact measurements of the IV curves for the samples used in our ARPES experiments. It is important to verify the relation of such IV curves to ones measured in a ``proper" 4-point contact geometry, where the voltage is measured across the sample itself, excluding any drop at the sample/contact interface. To accomplish this, we prepared a sample with contacts in standard 4-point geometry and simultaneously measured the current, voltage drop across the contacts (V$_c$) and sample (V$_s$). The measurements were conducted in ``constant" voltage mode, where we controlled the voltage across the contacts and the current was limited by the resistance of the sample and contacts. This prevents thermal runaway that commonly occurs in constant current measurements due to a rapid increase of the voltage and dissipated power as the sample transitions into the resistive state \cite{Kanigel}. Based on such data, we can measure the IV curves for 2-point contact and 4-point contact geometries at the same time under the same sample temperature, dissipated power and current conditions.  We apply a known voltage V$_c$ to the outer contacts and for each of its values we measure the current I and voltage on the inner electrodes V$_s$. The results are shown in Fig.~10.  The 2-point contact IV curve (I vs V$_c$) is plotted in red and is similar to data in the previous figures. The 4-point contact IV curve (I vs V$_s$) is plotted in blue. With double ended arrows we show examples of V$_c$ and V$_s$ pairs that were measured at the same time for the same current value. We first note that in the ``negative resistance" regime (right side of the current maximum) both curves are qualitatively similar. The current decreases with increasing voltage and reaches a minimum for higher voltage values. For our purposes, this is the most important point, since all the ARPES data shown in this paper were taken in this regime.
At very low voltages, there is of course a more significant difference between the two curves. In a 4-point contact measurement, the voltage is measured across a small section of the bridge and does not include the potential drop across the contacts. In the 2-point contact case, the voltage is measured across the whole bridge and contacts. Therefore, V$_s$ will remain zero up to a certain value of the current (signifying that the sample is in the zero resistance state). The value of V$_s$ will be lower than V$_c$ for the same current as this drop is measured over a shorter portion of the bridge and does not include the contact resistance. The current will initially increase linearly with V$_c$ voltage which reflects the resistance of the contacts and wires. This behavior is consistent with the insets of Figs.~2d and 3c, validating that part of the discussion of the IV curves.

\end{document}